\documentclass[aip,apl,preprint,amsmath,amssymb,showpacs,showkeys,floatfix]{revtex4-1}
\usepackage{graphicx}
\usepackage{color}
\usepackage{bm}
\begin{document}


\title{A new model for the recombination and radiative lifetime of trions and biexcitons in spherically shaped semiconductor nanocrystals}

\author{Mehmet Sahin}
\email{mehmet.sahin@agu.edu.tr; mehsahin@gmail.com}
\affiliation{Department of Material Science and Nanotechnology Engineering, Abdullah G\"{u}l University, Kayseri, Turkey}
\affiliation{Department of Physics, Faculty of Sciences, Selcuk University, 42075, Konya, Turkey}

\author{Fatih Ko\c{c}}
\affiliation{Department of Physics, Faculty of Sciences, Selcuk University, 42075, Konya, Turkey}

\begin{abstract}
In this letter, we propose a new model to determine the recombination oscillator strength of trions and biexcitons for bound and unbound cases in the effective mass approximation. The validity of our model has been confirmed by the radiative lifetime of the trion and biexciton in a spherical quantum dot. The results show that the model works sufficient accuracy in comparison with results of more complex methods such as quantum monte carlo techniques and atomistic calculations.
\end{abstract}

\maketitle

Semiconductor quantum dots (QDs) are very promising structures to confine basic carriers like electrons, holes, or excitons. Recent studies show that the exciton mechanisms in QDs play a very important role in revealing extraordinary optical properties and some quantum mechanical effects which can not be observed in bulk semiconductor materials.\cite{iva,meu,kim,fon,bro,ach} Quantum dots are currently the subject of intense research because of their controllable properties such as a size- and shape-tunable energy levels, lifetimes, high quantum yields, and chemical processability. These adjustable properties allow control of the dynamics of both single- and multi-exciton states by engineering the electronic and optical properties such as electron-hole wave function overlap, Coulomb interaction between the electron and hole, tunability between type-I and type-II localization regimes etc.\cite{kim,kor,bal,oro,pir,tyr} One of the best known optical properties is radiative decay of the excitons. In order to control the exciton radiative decay, one needs to tune the overlap between electrons and holes in QDs and this tuning is named as wave function engineering by Klimov et al.\cite{pir} The overlap is very important for oscillator strength and lifetime of the excitonic structures which can be tuned depending on the overlap. A number of studies have been reported on oscillator strength and lifetimes of neutral of charged exciton complexes.\cite{bac,com1,com2}

As is well known, respectively, the oscillator strengths of the exciton and biexciton structures in a quantum heterostructure are\cite{tak}

\begin{eqnarray}
\label{osc_X}
f_X \propto | < \psi_X |P_{cv}| 0 > |^2, \\ \nonumber
\\ \label{osc_XX}
f_{XX}\propto | < \psi_{XX} |P_{cv}| \psi_X > |^2,
\end{eqnarray}
where $|\psi_X>$ and $|\psi_{XX}>$ are the exciton and biexciton wave functions, respectively, and $P_{cv}$ is the momentum operator. Although these expressions are used for the absorption phenomena, the same forms are also used in the recombination process.

In this letter, we suggest an approximation to calculate the recombination oscillator strength of biexciton ($XX$), and negatively ($X^-$) and positively ($X^+$) charged excitons for bound and unbound cases in the effective mass approximation (EMA). In order to verify the model, we have calculated the radiative lifetime of the $X^-$, $X^+$ and $XX$ for bound and unbound cases in a spherical QD. We have shown that the recombination oscillator strengths and radiative lifetime of trions and biexcitons can be determined with sufficient accuracy by our simple model in comparison with results of more complex calculations.

In the electronic structure calculations, we have solved the Poisson-Schr\"{o}dinger equations self-consistently in the single band EMA to determine the energy levels and corresponding wave functions. Also, the quantum mechanical many-body interactions between the same kinds of particles have been taken into account in the local density approximation (LDA).

We consider a spherically symmetric QD structure. In the EMA and BenDaniel-Duke boundary conditions, the single particle Schr\"{o}dinger equations of a multi-exciton complex can be written as

\begin{equation}
\label{elham}
\left[-\frac{\hbar^2}{2}\vec{\nabla}_r \left(\frac{1}{m^*_e(r)}\vec{\nabla}_r \right) + V_e(r)-q_e \Phi_h + q_e\Phi_e + V_{xc}^{e-e}[\rho_e(r)]\right]R_{e}(r) = \varepsilon_{e} R_{e}(r),
\end{equation}
and
\begin{equation}
\label{hoham}
\left[-\frac{\hbar^2}{2}\vec{\nabla}_r \left(\frac{1}{m^*_h(r)}\vec{\nabla}_r \right) + V_h(r)-q_h \Phi_e + q_h\Phi_h + V_{xc}^{h-h}[\rho_h(r)] \right] R_{h}(r)=\varepsilon_{h} R_{h}(r),
\end{equation}
where $\hbar$ is the reduced Planck constant, $m^*_{e}(r)$ and $m^*_{h}(r)$ are the position dependent effective mass of the electron and hole, respectively, $V_e(r)$ is the electron confinement potential and $V_h(r)$ is the hole confinement potential, $q_e$ ($q_h$) is charge of the electron (hole), and $\Phi_e$ and $\Phi_h$ are the electrostatic Coulomb potentials of the electron and hole, respectively. The $V_{xc}[\rho(r)]$ potentials are the exchange-correlation potentials between the same kinds of particles, and $R_e(r)$ and $R_h(r)$ are the radial part of the electron and hole wave functions, respectively. $\varepsilon_{e}$ is the energy eigenvalue of the electron and similarly, $\varepsilon_{h}$ is the hole energy.

These two equations become coupled via attractive Coulomb terms ($q_e \Phi_h$ and $q_h \Phi_e$) and must be solved simultaneously with each other. At the same time, the self-consistency requirement in these calculations should be provided by the repulsive Coulomb potential terms ($q_e \Phi_e$ and $q_h \Phi_h$) in Eqs. (\ref{elham}) and (\ref{hoham}). In this way, all Coulomb effects on the energy eigenvalues and wave functions are taken into account. The electrostatic Coulomb potentials are calculated from the Poisson equations

\begin{eqnarray}
\label{pois}
\nonumber
\vec\nabla\kappa(r)\vec\nabla\Phi_{e} = \frac{q_e}{\varepsilon_0}\rho_{e}(r) \\ \nonumber
\\
\vec\nabla\kappa(r)\vec\nabla\Phi_{h} =-\frac{q_h}{\varepsilon_0}\rho_{h}(r),
\end{eqnarray}
where $\rho_{e}$ and $\rho_{h}$ are the density\cite{sah} of the electron and hole, respectively, $\varepsilon_0$ is dielectric permittivity of the vacuum and $\kappa(r)$ is the position dependent dielectric constant of the structure. These equations contain the image potential contributions due to surface polarization at the interfaces.

For the exchange-correlation potential in the trion and biexciton problem, the Perdew-Zunger\cite{perd} expression, which is a parametrization of the Monte Carlo results of Ceperley and Alder\cite{cep}, is employed. Also, this formulation contains the self-interaction correction.

The last three equations, Eqs.(\ref{elham}), (\ref{hoham}), and (\ref{pois}), are solved self-consistently by the full numeric matrix diagonalization technique. It should be noted that the repulsive Coulomb and exchange-correlation potential terms in both Eqs. (\ref{elham}) and (\ref{hoham}) must be reduced for $X$, since there are only one electron and one hole in the $X$. Similarly, these potential terms must be omitted in only Eq. (\ref{hoham}) for $X^-$ because of single hole and in only Eq. (\ref{elham}) for $X^+$ because of single electron.

The binding energy expressions of $X^-$, $X^+$, and $XX$ are given, respectively, as\cite{tsu}

\begin{eqnarray}
\label{bind}
E_{b}(X^-)=E_{X}^{tot}+\varepsilon_{e}(0)-E_{X^-}^{tot},
\nonumber\\
E_{b}(X^+)=E_{X}^{tot}+\varepsilon_{h}(0)-E_{X^+}^{tot},
\nonumber\\
E_b^{XX}=E_{XX}^{tot}-2E_X^{tot},
\end{eqnarray}
where $E_{X}^{tot}$ is single exciton total energy, $E_{X^-}^{tot}$ and $E_{X^+}^{tot}$ are the total energy of negatively and positively charged excitons, respectively, $E_{XX}^{tot}$ biexciton total energy and $\varepsilon_{e,h}(0)$ is isolated single electron (hole) energy in the QD.

As is well known, when a photon interacts with a QD, an electron-hole pair is formed in the QD. In this case, the single exciton oscillator strength, as is well known, is given by\cite{fon2}

\begin{equation}
\label{oscx}
f_X=\frac{E_p}{2E_X} \left|\int r^2 \mathrm{d}r R_e(r)R_h(r) \right|^2,
\end{equation}
where $E_p$ is the Kane energy, $E_{X}$ is the exciton transition energy, $R_e(r)$ and $R_h(r)$ are the radial part of electron and hole wave functions of the exciton, respectively. This equation can be easily used in both absorption and recombination processes of a single exciton.

\begin{figure}[t!]
\includegraphics[width=3in]{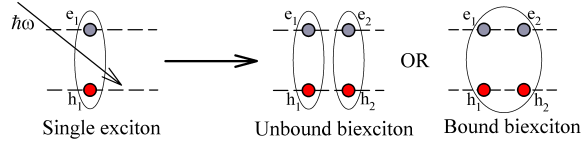}
\caption{Schematic representation of unbound and bound biexciton structure formation.}
\label{abs}
\end{figure}

\begin{figure}[t!]
\centering
\includegraphics[width=\columnwidth]{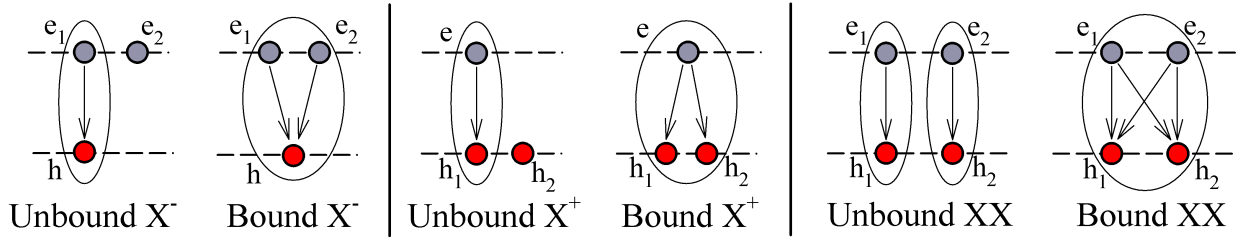}
\caption{Schematic representation of probable recombination processes in the unbound and bound $X^-$, $X^+$ and $XX$ structures, respectively.}
\label{recom}
\end{figure}

If a second photon is absorbed by the QD, this process results in a second exciton as illustrated in Fig.~\ref{abs}. This second exciton can occur in two possible situations, unbound or bound biexciton cases. In unbound biexciton case, the repulsive Coulomb interaction is predominant and this structure is considered as two-exciton isolated from each other. However, if the attractive Coulomb interaction is dominant, this complex is called as the bound biexciton and considered as a single particle. The oscillator strength of bound or unbound biexcitons is calculated by means of Eq. (\ref{osc_XX}). These forms of oscillator strengths are also used in recombination processes of any biexciton structure.\cite{fom,fel} This formulation is right for an absorption phenomena. On the other hand, we suggest that the oscillator strength of recombination is assumed as different in a trion or a biexciton structure. Because, in contrast to the absorption process, there are various situations in a recombination phenomena. As seen from the left panel of Fig.~\ref{recom}, the unbound $X^-$ structure is established  a single exciton and a one electron isolated from each other. In the unbound $X^-$ case, while one of the electrons (for example first electron) has the highest recombination probability with the hole, the recombination probability of second electron is approximately zero. This is similar to a single exciton recombination. However, in the bound $X^-$ case, all charges are bound with each other and considered as a single particle as seen the left panel of Fig.~\ref{recom}. Hence, both of the electrons have same recombination probabilities to the hole. Similar situations are valid for unbound and bound $X^+$, respectively, as seen in the middle panel of the figure. That is, while there is one recombination probability in the unbound trions as is in a single exciton, the recombination probability is two times higher in bound trions. Similar discussions can be made for the biexciton structure. In an unbound $XX$ case, while the recombination probability of first electron with first hole is the highest, the recombination probability of second electron with second hole is the highest. In this case, the recombination probability of an unbound $XX$ is two times larger than that of a single exciton. In a bound $XX$ case, the recombination probabilities of first electron with both holes are equivalent. Similarly, second electron has the same probability as that of the first one. As a result, the recombination probability of the bound $XX$ is two times greater than that of the unbound one.

In the light of all these discussions, we propose a different model to calculate the recombination oscillator strength of trions and biexcitons. In this model, the oscillator strength is to be computed using single particle wave functions of trions or biexcitons. The wave functions include all probable Coulomb interactions and the surface polarization effect because of the self-consistent electronic structure calculation procedure. The recombination oscillator strengths of trions and biexcitons are proposed as

\begin{equation}
\label{osc_t_b}
f_{(X^-,X^+,XX)} = A\frac{E_p}{2E_{(X^-,X^+,XX)}} \left|\int r^2 dr R_e(r) R_h(r) \right|^2,
\end{equation}
where $E_{X^-}$, $E_{X^+}$, $E_{XX}$ are the transition energies of negative or positive trion, or biexciton, respectively. $R_e(r)$ and $R_h(r)$ are the radial part of the electron and hole wave functions of the considered structure, $X^-, X^+$ or $XX$. Here, the $A$ is a recombination probability factor and $A\simeq2$ for bound and $A\simeq1$ for unbound trions. Similarly, the factor $A\simeq4$ for bound and $A\simeq2$ for unbound biexcitons.

The radiative lifetime of exciton is an important quantity for some device applications and therefore a number of studies on lifetime have been reported both theoretically and experimentally.\cite{iva,raj,gon,jha,nar,wan} In order to check the validity of our model, we calculate the radiative lifetime of bound and unbound trions and biexciton in a core/shell spherical QD. As is well known, the radiative lifetime is inversely proportional with the oscillator strength and it is defined as\cite{cal,ale}

\begin{equation}
\label{lifetime}
\tau = \frac{6\pi\varepsilon_0m_0c^3\hbar^2}{e^2n\beta_sE^2f},
\end{equation}
where $\varepsilon_0$ is the dielectric permittivity of the vacuum, $m_0$ is the free electron mass, $c$ is the light velocity, $e$ is the electronic charge, $f$ is the oscillator strength, $n$ is the refractive index, $E$ is the transition energy and $\beta_s$ is the screening factor.\cite{ale}

As a model structure, we use CdSe/CdS QD for type-I confinement regime. In this structure, both electron and hole are localized in the CdSe core region. The atomic units have been used throughout the calculations, $\hbar=m_0=e=1$. The material parameters are taken from Refs. \onlinecite{bro}. The effective exciton Bohr radius is 48.75 \AA\ and the effective exciton Rydberg energy is 15.9 meV.

\begin{figure}[t!]
\includegraphics[width=3in]{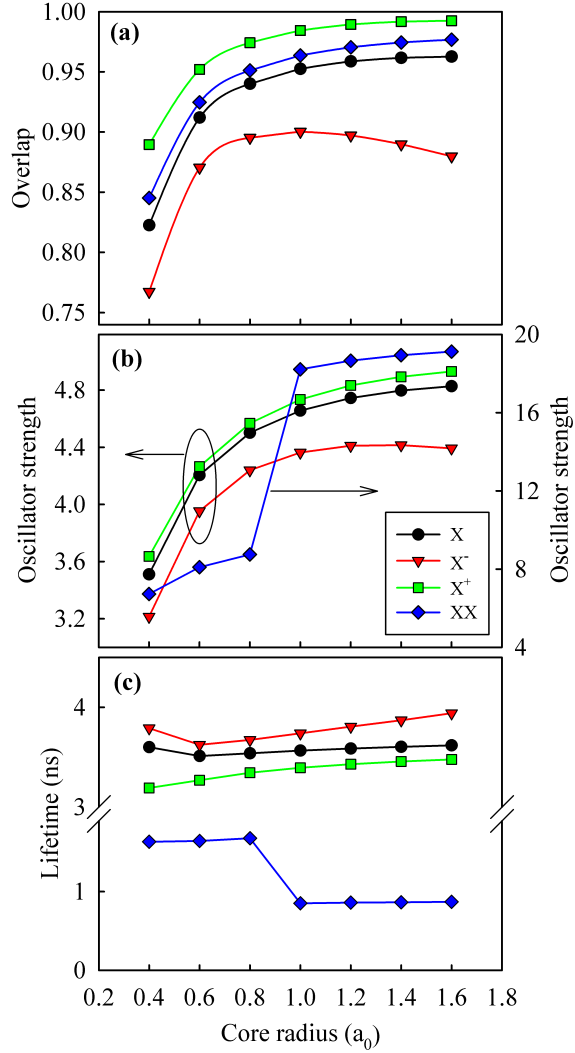}
\caption{The overlap integrals (a), oscillator strength (b) and lifetime (c) of the $X$, $X^-$, $X^+$ and $XX$ as a function of the core radius in CdSe/CdS QD.}
\label{osct1}
\end{figure}

In the CdSe/CdS QD, the calculations are performed as a function of the core radius for fixed shell thickness to 2 nm. In this structure, both of the electron and hole are confined to the CdSe core region. As can be seen from Eqs. (\ref{oscx}) and (\ref{osc_t_b}), the oscillator strength expressions are directly proportional to the electron-hole overlap integral. Figure \ref{osct1} (a) shows the overlap integral of the electron-hole wave functions for $X$, $X^-$, $X^+$, and $XX$. As is well known, in perturbation calculations, the overlap integral of the wave functions are same for the considered complexes, since there is no perturbative correction on the wave functions. Because all Coulombic effects in the electronic structure calculations are taken into account, the overlap integrals are different from each other. We conclude that the effects of the attractive and repulsive Coulomb potentials on the wave functions have drastically changed the overlap characteristic and hence these effects on the wave functions must be taken into account for more realistic calculations. The overlaps are small for all exciton complexes in small core radii and increase with increasing core radius except the $X^-$. The reason for this behavior in the $X^-$ can be explained with repulsive Coulomb energy between the electrons. Because the electrons are more energetic than the hole(s), their wave functions expand much more to the CdS shell region\cite{bro} and this process reduces the overlap of the wave functions. In the case of $X$, $X^+$ and $XX$, the kinetic energy term decrease with increasing core radii and the attractive Coulomb energy is a bit more dominant according to the $X^-$ and hence their overlap integrals increase.

The oscillator strength of $X$, $X^-$, $X^+$ and $XX$ are shown in Fig.~\ref{osct1} (b). The oscillator strengths of the $X^-$, and especially $X^+$, have approximately the same values as that of $X$ because the $X^-$ and $X^+$ are unbound in all core radii. When the core radius increases, the oscillator strength of $X^-$ decreases, while the oscillator strengths of $X$ and $X^+$ increase slightly. The $XX$ structure is unbound until the core radius is approximately equal to 4 nm. Therefore the recombination probability factor is $A\simeq2$. In further increasing of the core radii, the $XX$ becomes bound structure and therefore $A\simeq4$. The recombination oscillator strength of each exciton is $f_{X^-}/f_X\simeq0.92$ and $f_{X^+}/f_X\simeq1.02$. This relationship for biexciton is $f_{XX}/f_X\simeq1.92$ in unbound cases and $f_{XX}/f_X\simeq3.94$ in bound cases. These ratios increases with increasing core radius in both bound and unbound $XX$ cases.

The lifetime of the $X$, $X^-$, $X^+$ and $XX$, calculated by means of Eq.(\ref{lifetime}), is shown in Fig.~\ref{osct1} (c) as a function of the core radius. As seen from the figure, the single exciton lifetime is approximately equal for all core radii. Similar trends have been reported both theoretically and experimentally by some authors \cite{nar,bon}. Some changes have been observed in the trions radiative lifetimes with the core radius, but these variations are not very evident especially in $X^+$. Although, in some theoretical studies \cite{raj,nar}, it has been reported that the trion lifetime is almost half that of the single exciton lifetime, in our results, the trions' lifetimes are approximately equal to the exciton lifetime since the trions are not bound. The our model predicts $\tau_{X^-}^{-1}\simeq\frac{0.95}{\tau_X}$ and $\tau_{X^+}^{-1}\simeq\frac{1.05}{\tau_X}$. As regards to the biexciton, its lifetime is approximately two times (i.e. $\tau_{XX}^{-1}=\frac{2.14}{\tau_X}$) shorter than that of single exciton in case of unbound biexciton. This ratio is $\tau_{XX}^{-1}=\frac{4.17}{\tau_X}$ for the bound biexciton structure. Similar results are found from quantum monte carlo and atomistic calculations \cite{bac,nar,nar1,wim}.

In conclusion, we have suggested a new model to determine the recombination oscillator strength for the bound and unbound excitons in QDs and provided a different perspective for their radiative lifetimes. The model has been tested in determining the radiative lifetimes of the  $X^-$, $X^+$ and $XX$ in a type-I spherical QD heterostructures and it is seen that the results are in a very good agreement with quantum monte carlo and atomistic calculations results. We have not compared our results with the results of first-order perturbation calculations in the EMA because, in that method, there is not taken into consideration the Coulombic effect corrections on the wave functions. In addition, the binding energy results are used a different manner except traditional one and as depending on the binding energy, we have decided the probability of recombination type. The presented model can be expanded to type-II QD strucutures or to more excitonic complexes than a biexciton, for example three or four excitons and so on. It should be noted that the success of the model is strongly dependent on taking into consideration of Coulomb effects on the wave functions in the electronic structure calculations.

This study was supported by Turkish Scientific and Technical Research Council (TUBITAK) with Project Number 109T729.

\end{document}